\documentclass{article}
\usepackage{spconf,amsmath,graphicx}
\usepackage{multirow}
\usepackage{booktabs}
\usepackage{xcolor}
\usepackage[vlined,linesnumbered,ruled]{algorithm2e}
\usepackage[noadjust]{cite}

\title{Simulating realistic speech overlaps improves multi-talker ASR}
\name{\begin{tabular}{c}Muqiao Yang$^{1\dagger}$, Naoyuki Kanda$^2$, Xiaofei Wang$^2$, Jian Wu$^2$, Sunit Sivasankaran$^2$, Zhuo Chen$^2$,\\Jinyu Li$^2$, Takuya Yoshioka$^2$\end{tabular}\thanks{$^\dagger$Work performed during internship at Microsoft.}}
\address{   
$^1$ Carnegie Mellon University, Pittsburgh, PA, USA\\
$^2$Microsoft, Redmond, WA, USA}
\begin{document}
\ninept
\maketitle
\begin{abstract}
Multi-talker automatic speech recognition (ASR) has been studied to generate transcriptions of natural conversation including overlapping speech of multiple speakers. Due to the difficulty in acquiring real conversation data with high-quality human transcriptions, a naïve simulation of multi-talker speech by randomly mixing multiple utterances was conventionally used for model training. In this work, we propose an improved technique to simulate multi-talker overlapping speech with realistic speech overlaps, where an arbitrary pattern of speech overlaps is represented by a sequence of discrete tokens. With this representation, speech overlapping patterns can be learned from real conversations based on a statistical language model, such as N-gram, which can be then used to generate multi-talker speech for training. In our experiments, multi-talker ASR models trained with the proposed method show consistent improvement on the word error rates across multiple datasets.
\end{abstract}
\begin{keywords}
Multi-talker automatic speech recognition, conversation analysis, data simulation
\end{keywords}

\newcommand{\ra}[1]{\renewcommand{\arraystretch}{#1}}

\makeatletter
\newcommand{\removelatexerror}{\let\@latex@error\@gobble}
\makeatother

\section{Introduction}
\vspace{-.5em}
\label{sec:intro}

Automatic speech recognition (ASR) plays a crucial role in human-machine interactions and human conversation analyses. Even though there has been a significant progress of ASR technology in recent decades \cite{seide2011conversational, hinton2012deep, qian2016very}, it is still challenging to transcribe natural conversation because of the complicated acoustic and linguistic properties. Especially, natural conversation contains a considerable amount of speech overlaps \cite{ccetin2006analysis}, which significantly hurt the accuracy of conventional ASR systems designed for single-talker speech~\cite{chen2020continuous,raj2020integration}. 
%there still exist many challenging scenarios for ASR including overlapped speech, far-field condition \cite{chang2020end, haeb2020far}, etc. 
To overcome the limitation, 
 multi-talker ASR that generates transcriptions of multiple speakers has been studied~\cite{yu2017recognizing,seki2018purely,chang2019end,chang2019mimo,tripathi2020end,kanda2020serialized,lu2021streaming,sklyar2021streaming,kanda2022streaming,kanda2022vararray}.
 To train such multi-talker ASR models in high quality, it is crucial to feed the model a sufficient amount of multi-talker speech samples with accurate transcriptions. 

%Usually, it is often difficult to collect real conversation data with high quality transcriptions.
% Compared to single-talker ASR which aims to decode the transcription of a single speaker with the presence of background noise, this task is more difficult and requires a large amount of annotated speech data to get a reasonable performance.

Most prior works used the simulated
multi-talker speech generated from single-talker ASR training data~\cite{yu2017recognizing,seki2018purely,chang2019end,chang2019mimo,tripathi2020end,kanda2020serialized,lu2021streaming,sklyar2021streaming,kanda2022streaming,kanda2022vararray} since it is expensive to collect a large number of real conversations with high quality transcriptions. 
For example, 
Seki et al.~\cite{seki2018purely} mixed two single-talker audio samples 
where a shorter sample is mixed with a longer one with a random delay 
such that the shorter sample is fully overlapped with the longer sample.
Kanda et al.~\cite{kanda2020serialized} mixed $S$ single-talker samples with random delays to
simulate partially overlapping speech, where $S$ was uniformly sampled from one to five. 
While there were minor differences in each work,
all prior works naively mixed single-talker speech samples with random delays,
which incurs
 unnatural speech overlapping pattern in the training data.
 On the other hand, it was suggested that 
 a large portion of the degradation of word error rates (WER) was caused by the insufficient training of speech overlapping patterns \cite{kanda2021large}. 
%of these mixture methods are based on a random simulation mechanism, while its simulated conversations may have some limitations and flaws. One of the limitations is that the overlap pattern of the simulated conversations might be quite different from realistic conversations. 
% The previous study \cite{kanda2021large} has shown that overlap pattern is an important factor to reduce the WER (word error rate) in multi-talker ASR. 
Recently, a few trials were made to simulate more realistic multi-talker audio in the context of speaker diarization. 
Landini et al. \cite{landini2022simulated} used  statistics about frequencies and lengths
of pauses and overlaps. 
Yamashita et al. \cite{yamashita2022improving} used the transition probabilities between different overlap patterns.
%to determine which transition type to be used in the simulation. 
Although 
these works demonstrated superior performance for speaker diarization accuracy, 
such an approach has not been studied for multi-talker ASR. % where the 
In addition, these methods are not directly applicable because they are assuming speaker information available for every speech sample while the ASR training data is often anonymized by excluding speaker identifiable information~\cite{kanda2021large}.

In this work, we propose a novel multi-talker speech data simulation method for multi-talker ASR training.
We first represent
an arbitrary pattern of speech overlaps by a sequence of discrete tokens.
With this representation, speech overlapping patterns can be learned from real conversations
based on 
 a statistical language model (SLM), such as N-gram.
 The learned overlapping pattern can be then used to simulate multi-talker speech with realistic speech overlaps.
% Unlike prior methods which rely on the statistics of limited aspects of speech overlaps~\cite{landini2022simulated,yamashita2022improving},
% our method can directly represent arbitrary pattern of speech overlaps.
% , to better model the patterns from real corpora and help simulate statistically realistic conversations in multi-talker ASR as a generation model, as well as getting rid of the requirement of speaker information. 
We demonstrate the accuracy of multi-talker ASR can be consistently improved based on the proposed multi-talker simulation data 
across multiple real meeting evaluation sets.
We also show that the accuracy is further improved by combining different types of simulation data as well as a small amount of real conversations.

\section{Related Works}
\vspace{-.5em}
\label{sec:related_work}
% \subsection{Multi-talker ASR with token-level serialized output training}
\subsection{Token-level serialized output training for multi-talker ASR}
\vspace{-.5em}

The token-level serialized output training (t-SOT) 
was proposed to generate transcriptions of multi-talker overlapping speech 
in a streaming fashion \cite{kanda2022streaming}. 
In the t-SOT framework, the maximum number of active speakers at the same time is assumed to be $M$.
In our work, we used the t-SOT with $M=2$ for its simplicity, thus explaining the t-SOT with this condition.
With t-SOT, 
the transcriptions of multiple speakers are 
serialized into a single sequence of 
recognition tokens (e.g., words, subwords, etc.) 
by sorting them in chronological order. 
A special token $\langle \mathrm{cc}\rangle$, indicating a change of virtual output
channels, 
is inserted between two adjacent words spoken by different speakers. 
A streaming end-to-end ASR model \cite{li2022recent} is trained 
based on pairs of such serialized transcriptions and the corresponding audio samples.
During inference, 
an output sequence including 
$\langle \mathrm{cc}\rangle$ is generated
by the ASR model in a streaming fashion, 
which is then converted to separate transcriptions 
based on the recognition of the $\langle \mathrm{cc}\rangle$. 
The t-SOT model showed superior performance compared to prior multi-talker ASR models
even with streaming processing.
See \cite{kanda2022streaming} for more details.

\subsection{Multi-talker speech simulation based on random mixtures}
\vspace{-.5em}
% \subsection{Random simulation}
% Due to the limited amount of real conversation data with high quality human transcription,
Prior multi-talker ASR studies  used 
a simple simulation technique to generate training data by mixing single-talker speech signals 
with random delays~\cite{yu2017recognizing,seki2018purely,chang2019end,chang2019mimo,tripathi2020end,kanda2020serialized,lu2021streaming,sklyar2021streaming,kanda2022streaming,kanda2022vararray}.
% are simulating their multi-talker conversations from single-talker speech with a random simulation mechanism. 
In this work, as a baseline technique, we adopt the procedure used in
 \cite{kanda2022streaming} where multiple single-talker audio samples are mixed with random delay 
  while limiting the maximum number of active speakers at the same time to up to two.
Let $S_0$ denote a
 single-talker ASR training set.
For each training data simulation, 
the number of audio signals $\tilde{K}$ to be mixed is first uniformly sampled from 
1 to $K$, where $K$ is the maximum number of utterances to be mixed.
Then, $\tilde{K}$ utterances are randomly sampled from $S_0$.
Finally, 2nd to $\tilde{K}$-th utterances are iteratively mixed with the 1st utterance after prepending random silence signals.
The duration of the silence prepended for $k$-th utterance is
sampled from $\mathcal{U}(\mathrm{end2}(a_{k-1}), \mathrm{len}(a_{k-1}))$, where
$\mathcal{U}(\alpha,\beta)$ is the uniform distribution function in the value range of $[\alpha,\beta)$,
$a_{k-1}$ is the mixed audio after mixing $(k-1)$-th utterance,
$\mathrm{len}(a)$ is a function to return the duration of audio $a$,
and
 $\mathrm{end2}(a)$ is a function to return
the end time of the penultimate utterance in $a$.

% under the constraints that (i) the generated training sample contains the overlapping speech of up to two speakers at the same time and (ii) each training sample has at least one overlapping region with other training sample.
% This can be achieved by sampling the silence duration 
% from $\mathrm{Uniform}(t_)$,
% which are achieved by using the procedure proposed in \cite{}.
% For each pair of adjacent utterances $(u_1, u_2)$ in the chronological order, it will randomly sample the delay value $d\sim \text{Uniform}(0, \text{len}(u_1))$, where $\text{len}(u_1)$ is the duration of the first utterance $u_1$. After that, $u_1$ and $u_2$ can be simulated together by adding a delay of $d$ before the second utterance $u_2$. By iteratively performing such kind of random simulation, it would be able to generate infinite variations of training samples.

% Specifically, t-SOT is generating their serialized multi-speaker transcriptions by outputting a sequence of words from multiple speakers with special separator symbol $\langle cc\rangle$.

\section{Proposed Multi-talker speech simulation}
\vspace{-.5em}
\label{sec:method}

\subsection{Overview}
\label{subsec:discretize}
\vspace{-.5em}

In this work,
we propose to simulate multi-talker speech 
based on the statistics of speech overlaps in real conversation.
% for better multi-talker ASR training.
% Such a simulated multi-talker speech is used for training multi-talker ASR models with improved accuracy.
The overall workflow of our proposed method is as follows.
(1) Convert time- and speaker-annotated transcription of real conversation into a sequence of discrete tokens, with each representing the status of speaker overlap within a short time- or word-based unit. (2) Train SLM $\mathcal{M}$ using the discrete token sequence. (3) Simulate multi-talker speech based on single-talker ASR training data $S_0$ and the SLM $\mathcal{M}$. (4) Train multi-talker ASR by using the simulated multi-talker speech.

The core of our proposed method lies in converting the time- and speaker-annotated transcription 
 into a discrete token sequence that represents the speaker overlapping pattern.
% There are multiple possible ways for such conversion,
% to convert speech overlapping pattern into
% a sequence of discrete tokens.
In this paper, 
we propose two algorithms for such conversion, named time-based discretization and
word-based discretization, which
are explained in Section \ref{subsec:time-based} and Section \ref{subsec:word-based}, respectively.
Note that our algorithms use the notion of ``virtual channel'' proposed in t-SOT~\cite{kanda2022streaming}. %, and can be easily extended for the case of $M>2$ by introducing more virtual channels.
In the following explanation, we assume two virtual channels (i.e. $M=2$) for simplicity and consistency with the t-SOT.
%$M=2$, i.e., the maximum number of concurrently active speakers is up to two for simplicity. %, for both real conversation and
Our algorithm can be easily extended for $M>2$ by assuming more virtual channels.
% simulated data 
%in order to be consistent with t-SOT based ASR.

% This section describes our proposed multi-talker speech simulation method.
% We first explain the key idea to represent an arbitrary pattern of speech overlaps based
% on a sequence of discrete tokens in Section \ref{subsec:discretize}. 
% We then explain how we can generate multi-talker audio samples
% based on the discretized overlapping pattern in Section \ref{subsec:generate}. 
% In the following explanation, we assume $M=2$ to be consistent with t-SOT based ASR.
%We will use t-SOT as the backbone of multi-talker modeling. To be consistent with t-SOT, we restrict the maximum number of concurrent speakers to 2 in all of our simulations. 

\SetKwComment{Comment}{// }

\newcommand\mycommfont[1]{\footnotesize\ttfamily\textcolor{black}{#1}}
\SetCommentSty{mycommfont}

% \begin{algorithm}[h]
% \DontPrintSemicolon
%   \KwSty{Input: } Set of utterances $\mathcal{U}$, N-gram model $\mathcal{M}$, max. utt. to be mixed in random simulation $N_{max}$, audio pool size $K$ \\
%   \KwSty{Output: } Simulated multi-talker utterances $\{x\}$
%   
% \For{$i\gets1$ \KwTo $K$}{Randomly sample one audio from $\mathcal{U}$ and push it to the audio pool with its word length and time length information}
% 
% \While{True}{
% Randomly generate a sequence of states $\mathcal{S} = \{s\}_{i=1}^n $ from $\mathcal{M}$ where $s \in \{00, 01, 10, 11\}$  \ \tcp*{each state represents either word or time}
% 
% Decide the simulation type $P_s \in \{\textit{random, word, time}\}$ with some probability
% 
% \eIf{$P_s$ == random}{Sample $N\sim\text{Uniform}(1, N_{max})$ and randomly take $N$ utterances from the pool \\
% Sample $D_r = \{d\}_{i=2}^N$ where $d_i\sim\text{Uniform}(0, \text{Len}(d_{i-1}$)) \\}{\eIf{$P_s$ == word}{
% Extract overlap pattern: Time length information $L_w$ and delay information $D_w$ from $\mathcal{S}$}{Extract overlap pattern: Word length information $L_t$ and delay information $D_t$ from $\mathcal{S}$} Pop utterances with the closest length to the length information $L$ from the pool}
% 
% Simulate $x$ by using the length information $L$ and delay information $D$
% 
% Push new utterances from $\mathcal{U}$ to the pool until its size is $K$ again
% 
% }
% \caption{N-gram simulation}
% \end{algorithm}

 \begin{figure}[t]
\removelatexerror
\begin{algorithm}[H]
%\begin{algorithm}[h]
\footnotesize
% \scriptsize
\DontPrintSemicolon
  \KwSty{Input: } A speech sample with a time- and speaker-annotated transcription $\mathcal{T}=(w_i, b_i, e_i, s_i)_{i=0}^{N-1}$ where $w_i$ is $i$-th word, 
  $b_i, e_i, s_i$ is the begin time, end time, and speaker index of $w_i$, respectively.
  Duration of one discretization unit $d$.\\
  \KwSty{Output: } discretized speaker overlapping pattern $\mathcal{X}$
  
Sort $\mathcal{T}$ in ascending order of $e_i$.
 
Current channel index $c\leftarrow 0$. 
 
Initialize a list of channel indices $\mathcal{C}\leftarrow[0]$. 

\For{$i\gets1$ \KwTo $N-1$}{
    \If{$s_i \neq s_{i-1}$}{$c\leftarrow 1-c$}
    
    Append $c$ to $\mathcal{C}$
}

\eIf{(Time-based discretization)}{

  \For{$t\gets0$ \KwTo $\lfloor e_{N-1}/d\rfloor$}{
  
      $q \leftarrow {\tiny\begin{bmatrix} 0 \\0 \end{bmatrix}}$
  
      % $b^{\rm slot} \leftarrow d \cdot t$
      
      % $e^{\rm slot} \leftarrow d \cdot (t+1)$
      
      \For{$i\gets0$ \KwTo $N-1$}{
          \If{$[b_i, e_i]$ is overlapped with $[d \cdot t, d \cdot (t+1)]$}{$q[\mathcal{C}[i]] \leftarrow 1$}
      }
      $x \leftarrow q[0] + 2\cdot q[1]$ \tcp*[r]{\it Encode to discrete token} 
      
      Append $x$ to $\mathcal{X}$
  }
}{ \vspace{-3mm} \tcp*[l]{\it (Word-based discretization)} 

  \For{$i\gets0$ \KwTo $N-1$}{
  
      $q \leftarrow {\tiny\begin{bmatrix} 0 \\0 \end{bmatrix}}$
      
      \For{$j\gets0$ \KwTo $N-1$}{
          % \If{$b_i < e_j \And b_j < e_i$}{$u[\mathcal{C}[i]] \leftarrow 1$}
          \If{$[b_j, e_j]$ is overlapped with $[b_i, e_i]$}{$q[\mathcal{C}[j]] \leftarrow 1$}
      }
      $x \leftarrow q[0] + 2\cdot q[1]$ \tcp*[r]{\it Encode to discrete token}
      
      Append $x$ to $\mathcal{X}$
  }
}% \Else{}
Return $\mathcal{X}$
\caption{\footnotesize Discretization of speech overlapping pattern with up to two concurrent utterances}
\label{algo:discretization}
\end{algorithm}
\vspace{-6mm}
\end{figure}

\begin{figure}[t]
\removelatexerror
\begin{algorithm}[H]
%\begin{algorithm}[h]
\footnotesize
\DontPrintSemicolon
  \KwSty{Input: } Single-talker ASR training data set $S_0$. 
  SLM learned from the discretized speaker overlapping pattern $\mathcal{M}$.  Duration of one discretization unit $d$. \\
  \KwSty{Output: } Simulated multi-talker speech sample $a$
  
Sample a discretized token sequence $\tilde{\mathcal{X}}$ from $\mathcal{M}$

$\tilde{\mathcal{Q}}\leftarrow {\rm Decode}(\tilde{\mathcal{X}})$
\tcp*[r]{\it 
$0\to{\tiny\begin{bmatrix}0\\0\end{bmatrix}}$,
$1\to{\tiny\begin{bmatrix}1\\0\end{bmatrix}}$,
$2\to{\tiny\begin{bmatrix}0\\1\end{bmatrix}}$,
$3\to{\tiny\begin{bmatrix}1\\1\end{bmatrix}}$} 

Initialize mixed audio $a \leftarrow []$

End times of each word $D \leftarrow []$  \tcp*[r]{\it Only for word-based disc.} 

\For{($i^b, i^e)\gets {\rm ConsecutiveOne}(\tilde{\mathcal{Q}})$}{
\eIf{(Time-based discretization)}{
    $d^{\rm min} \leftarrow d \cdot (i^e - i^b)$ % \tcp*[r]{\it Minimum target duration} 
    
    $d^{\rm max} \leftarrow d \cdot (i^e - i^b + 1)$ % \tcp*[r]{\it Maximum  target duration} 
    
    $u \leftarrow \mathrm{DurationNearestSample}(d^{\rm min}, d^{\rm max}, S_0)$ 
    
    $\gamma \leftarrow d \cdot i^b + \mathcal{U}(0, d^{\rm max}-\mathrm{len}(u))$ \tcp*[r]{\it Silence duration} 
    
}{ \vspace{-3mm} \tcp*[l]{\it (Word-based discretization)} 

    $\mathcal{I}=[i_0,...,i_{M-1}]\gets \mathrm{WordIndices}(i^b, i^e, \tilde{\mathcal{Q}})$
    
    % $w \leftarrow \mathrm{len}(\mathcal{I})$ \tcp*[r]{\it Target word count} 
    $u \leftarrow \mathrm{WordCountNearestSample}(\mathrm{len}(\mathcal{I}), S_0)$
    
    $\gamma \leftarrow D[i_0-2] + \mathcal{U}(0, D[i_0 - 1] - D[i_0 - 2])$
   
    \For{$j\gets0$ \KwTo $M-1$}{ 
    %\For{$i\gets \mathcal{I}$}{
      $D[i_j] \leftarrow$ end time of $j$-th word in $u$
    }

}
    $a \leftarrow a + [sil(\gamma), u]$
}

Return $a$
\caption{\footnotesize Multi-talker speech generation.}
\label{algo:generation}
\end{algorithm}
\vspace{-5mm}
\end{figure}

\subsection{Time-based discretization}
\label{subsec:time-based}
\vspace{-.5em}

%The motivation behind the proposed method is that the various aspects of randomness in the random simulation may make the simulated conversations far from realistic ones. Instead, we would expect the simulated conversations to be as similar as possible to real conversations, in terms of the speech overlapping patterns. To achieve this objective, we can model the overlapping pattern with an SLM in two different perspectives, time-based discretization and word-based discretization. The overall discretization process is described in Algorithm \ref{algo:discretization}. 

% \subsubsection{Time-based discretization}
% \vspace{-.5em}

In this method, 
the speech overlapping pattern 
is represented by a sequence of discrete tokens,
where each token indicates the speech overlapping status of
a $d$-sec time window. 
The algorithm is depicted in line 1--17, 26 of Algorithm \ref{algo:discretization}. 
% Note that we are training the SLM with corpora where the speaker information is available, so we are able to access the speaker index $s$ during the discretization process. 
% Given a short transcribed conversational segment with each word having timing and speaker markers, we first sort $\mathcal{T}$ based on the end time of each word.
% We then assign each word a ``virtual'' channel index $c\in\{0,1\}$. 
Suppose we have a time- and speaker-annotated transcription $\mathcal{T}$ of a
small amount of real conversation data.
% % of $N$ words, 
We first sort $\mathcal{T}$ based on the end time of each word.
We then assign each word
a ``virtual'' channel index $c\in\{0,1\}$.
Here,
 $c=0$ is always assigned to the first (i.e 0-th) word (line 4, 5). 
For words thereafter, the value of $c$ is changed 
when the adjacent words
are spoken by different speakers, which is the sign of 
either speech overlap or speaker turn (lines 7--9).
% The assigned channel index is stored in the list $\mathcal{C}$ (line 9).
% based on the speaker turn when we,
% initialize a list of channel indices $\mathcal{C}$ by the speaker index information $\{s\}_{i=1}^N$. 
Once the list $\mathcal{C}$ of the channel indices is prepared, the transcription is converted
to a sequence $\mathcal{X}$ % by summing up the virtual channels. $\mathcal{X}$ then contains 
of discrete tokens $x\in\{0,1,2,3\}$, 
which indicates the speech activity of each virtual channel for a short $d$-sec time window (lines 10--17).
More specifically, $x=0$ (or $q={\tiny\begin{bmatrix}0\\0\end{bmatrix}}$) indicates no speech activity is observed during the $d$-sec time window.
On the other hand, $x=1, 2, 3$ (or $q={\tiny\begin{bmatrix}1\\0\end{bmatrix},\begin{bmatrix}0\\1\end{bmatrix},\begin{bmatrix}1\\1\end{bmatrix}}$) indicates that 
the speech activity is observed 
in 0-th channel, 1-st channel, 
or both channels
during the $d$-sec time window.

% $\tiny\begin{bmatrix} 0 \\1 \end{bmatrix}$

% The entire length of the overlapping pattern is determined by its total duration $e_N$ divided by $d$, rounding up to integers. Then, for each word in this speech sample, we will represent whether it is overlapped with each time unit $[d \cdot t, d \cdot (t+1)]$ by a binary value. Given the corresponding channel index of each word, we will use 1 to represent overlapping and 0 as non-overlapping. As the final step, a two-digit binary number $x$ (valued from 0 to 3) is used to represent the overlapping state of one specific time unit, where each digit represents the state of the corresponding channel. Such a process is iterated for all the time units to create the final discretized speaker overlapping pattern $\mathcal{X}$.

% Once we convert the transcriptions of real conversation into a sequence of discrete tokens,
SLM $\mathcal{M}$ can be trained based on the derived sequence of discrete tokens, and then used to yield speech overlapping patterns for simulating more realistic multi-talker speech.
The algorithm of generating a multi-talker speech sample 
is described in lines 1--12, 19, 20 of Algorithm \ref{algo:generation} and depicted as follows. 
% Suppose we have an SLM $\mathcal{M}$ that is trained from
% a time-discretized overlapping pattern of real conversation based on Algorithm \ref{algo:discretization}. 
We first sample a discretized token sequence $\tilde{\mathcal{X}}$ from SLM $\mathcal{M}$ (line 3) based on the time-based discretization. 
We then apply $\rm Decode$ function that converts $\tilde{\mathcal{X}}$ into a sequence $\tilde{\mathcal{Q}}$,  consisting of a list of binary pairs $q\in\{{\tiny\begin{bmatrix}0\\0\end{bmatrix},\begin{bmatrix}1\\0\end{bmatrix},\begin{bmatrix}0\\1\end{bmatrix},\begin{bmatrix}1\\1\end{bmatrix}}\}$, where 1 in $c$-th element of $q$ indicates the existence of speech in the $c$-th channel (line 4).  
% After initializing theing audio $a$ is initialized  (line 5), 
Next, a target mixed audio $a$
% , whose final duration is based on a $\rm ConsectiveOne$ function, 
is initialized with empty (line 5).
We then call $\rm ConsectiveOne$ to extract
the beginning index $i^b$ and ending index $i^e$ of the consecutive sequence of $1$ in either 0-th or 1-st channel of $\tilde{\mathcal{Q}}$, in chronological order of $i^b$ (line 7). For example, when $\tilde{\mathcal{Q}}=
[{\tiny\begin{bmatrix}1\\0\end{bmatrix},\begin{bmatrix}1\\0\end{bmatrix},\begin{bmatrix}1\\1\end{bmatrix},\begin{bmatrix}1\\1\end{bmatrix},\begin{bmatrix}0\\1\end{bmatrix}}]$, $\rm ConsectiveOne$ returns $i^b=0, i^e=3$ for the first iteration,
and returns $i^b=2, i^e=4$ for the second iteration.
Based on $i^b$ and $i^e$, 
we sample a speech segment $u$ from source ASR training data $S_0$, with the duration within or nearest to
the expected interval [$d^{\rm min}$, $d^{\rm max}$]  (lines 9--11).\footnote{To save computation, we created a pool of 10K utterances 
from $S_0$, and used it to sample $u$.
When we created the pool, utterances in $S_0$ were short-segmented at the point of
silence longer than 0.5 sec.
We randomly sampled $u$ among samples whose duration
is within [$d^{\rm min}, d^{\rm max}$]. If there was no such sample, we sample
an utterance having closest duration to $(d^{\rm min}, d^{\rm max})/2$.}
A duration $\gamma$ of the silence prepended to $u$ is randomly determined such that $u$ is mixed to the audio $a$ with the expected position from $i^b$ (line 12, 19). 
These procedures are repeated to form the mixed audio $a$, which is
 returned as the training sample for multi-talker ASR (line 20).

\subsection{Word-based discretization}
\vspace{-.5em}
\label{subsec:word-based}

The word-based discretization algorithm differs from the time-based
one in a sense that the discretized unit representing the overlapping status
is based on ``word'' rather than a $d$-sec time window.
The algorithm of converting $\mathcal{T}$ into the discretized speaker overlapping pattern
$\mathcal{X}$ is depicted in lines 1--9, 18--26 of Algorithm \ref{algo:discretization}. 
The word-based discretization shares the same procedure with the time-based discretization for the first 9 steps, and then performs discretization for each word in $\mathcal{T}$, where
$x$ represents the overlapping status of speech overlaps of each token (lines 18--25).
Note that $x$ cannot be 0 (or $q$ cannot be $\tiny\begin{bmatrix}0\\0\end{bmatrix}$)
in this algorithm. Fundamentally, word-based discretization is assumed to be applied
to a short segment that does not include long-silence signals.
This can be achieved by segmenting the transcription 
based on the existence of long silence.
% with voice activity detection to guarantee there is no long silence signals

% the overlapping pattern with the total number of words in the speech sample, i.e. $N$. For each word $i$, we will iteratively look through all other words to see if it is overlapped with any other word $[b_j, e_j]$, and similarly, we will use a binary value to indicate the overlapping state. In the end, the discretized speaker overlapping pattern $\mathcal{X}$ of length $N$ is produced to train the SLM. Note that different from time-based discretization, word-based discretization gets rid of one additional hyperparameter, the discretization time unit $d$.

% \subsection{Generating multi-talker audio samples}
% \label{subsec:generate}
% \vspace{-.5em}
% \subsubsection{Time-based discretization}
% \vspace{-.5em}

% The second stage of the proposed method is to generate multi-talker audio samples from the trained SLM. Since we have introduced two different discretization mechanisms to train the SLM, the generation processes of time-based and word-based discretization are also different. 

% \subsubsection{Word-based discretization}
% \vspace{-.5em}

The multi-speaker speech generation algorithm with word-based discretization is introduced in Algorithm \ref{algo:generation}.
%, starting from an SLM $\mathcal{M}^{\rm word}$ trained from word-based discretized patterns in Algorithm \ref{algo:discretization}. The generation process is mostly similar with the time-based one by first decoding the discretized token sequence $\tilde{\mathcal{X}}$ into the binary value seuqnce $\hat{\mathcal{X}}$ (line 4), and then iteratively looking for consecutive $1$s with the $\rm ConsectiveOne$ function in the discretized token sequence $\hat{\mathcal{X}}$ (line 7). 
The first 7 steps are the same as the time-based discretization algorithm
except that we create a buffer $D$ to keep the end time of each word of utterances being mixed (line 6).
The buffer $D$ is necessary to determine the duration $\gamma$ of silence
when we mix $u$ to $a$ with targeted overlaps (lines 16--18).\footnote{We assume $D$ will return 0 for an undefined index.}
In line 14, we first apply $\rm WordIndices$ function, 
which returns a list $\mathcal{I}$ of indices indicating where each word should be allocated.
This function is necessary because
 around half of the consecutive sequence of $\tiny\begin{bmatrix}1\\1\end{bmatrix}$  represents the words in the other channel. 
For example, given $\tilde{\mathcal{Q}}=
[{\tiny\begin{bmatrix}1\\0\end{bmatrix},\begin{bmatrix}1\\0\end{bmatrix},\begin{bmatrix}1\\1\end{bmatrix},\begin{bmatrix}1\\1\end{bmatrix},\begin{bmatrix}0\\1\end{bmatrix}}]$,
$\rm ConsectiveOne$ first returns $i^b=0, i^e=3$,
and $\rm WordIndices$ returns $[0,1,2]$.
For the second iteration, 
$\rm ConsectiveOne$ returns $i^b=2, i^e=4$,
and $\rm WordIndices$ returns $[3,4]$.
We then sample $u$ whose number of words are closest to the length of $\mathcal{I}$ (line 15).\footnote{We used a pool of 10K utterances 
from $S_0$,
and randomly sampled $u$ among samples whose number of words
is $\mathrm{len}(\mathcal{I})$.
If there was no such sample, we sample
an utterance having the closest word count to $\mathrm{len}(\mathcal{I})$.}
We mix $u$ to $a$ with a silence signal whose duration $\gamma$ is determined such that
the word in $u$ is overlapping with $a$ as represented by $\mathcal{Q}$ (lines 16, 19).

\begin{table}[t]
    \vspace{-3mm}
    \caption{Datasets used in the experiments.}
    % \caption{Real meeting datasets used in the experiments. Besides these datasets, we also used 75K hours of single-talker ASR training data for the mulit-talkedata simulation. }
    \label{tab:dataset}
    \ra{0.9}
    \tabcolsep = 0.9mm
    \footnotesize
    \centering
    \begin{tabular}{c|ccc|ccc|c|c}
    \toprule
    % \multirow{3}{*}{Simulation Method} &
    & 
    \multicolumn{3}{c|}{AMI} &
    \multicolumn{3}{c|}{ICSI} &
    \multirow{2}{*}{MS$^{\rm mtg}$}  % \\
      & \multirow{2}{*}{MS$^{\rm asr}$}  \\
     & 
    \multirow{1}{*}{train} &
    \multirow{1}{*}{dev} &
    \multirow{1}{*}{eval} &
    \multirow{1}{*}{train} & 
    \multirow{1}{*}{dev} &
     \multirow{1}{*}{eval} &
     % &
     \\
     \midrule
     Size (hr) & 80.2 & 9.7 & 9.1 & 66.6   & 2.3 & 2.8 & 15.5 & 75K   \\ 
     \# meetings & 135 & 18 & 16 & 70 & 2 & 3 & 33 & -   \\ 
     \# speakers / meeting & 3--5 & 4 & 3--4 & 3--10 & 6--7 & 7 & 2--19  & -   \\ 
     \midrule
     Overlap SLM training & $\surd$ & - &  -& - & - & - & - & - \\
     % Data Simulation   &  -      &  - & - &  - & - & - &  - & $\surd$ \\
     t-SOT pre-training  & - & - & - & - & - & - & - & $\surd$ \\
     t-SOT fine-tuning  & $\surd$ & - & - & $\surd$ & - & - & - & $\surd$ \\
     Evaluation   &  -      &$\surd$ &$\surd$ & - & $\surd$ & $\surd$ & $\surd$ & - \\
    \bottomrule
    \end{tabular}
    \vspace{-6mm}
\end{table}

\section{Experiments}
\vspace{-.5em}
\label{sec:exp}

% \textcolor{red}{MEMO by Naoyuki: Do we want to add the result with enhanced audio such that the traial failed? As a comparison to the data filtering. [Comment by TY] I don't think this is necessary. In my view, in order to draw reliable conclusions, we'll need to do more experiments, e.g., adding distortion/noise again to the mixtures of the cleaned signals or testing with IHM-mix.}

% \textcolor{red}{MEMO by Naoyuki: We want to show 1spk ASR result somewhere}.

\subsection{Evaluation data}
\label{subsec:data}
\vspace{-.5em}

%All of our experiments are fine-tuning a pre-trained model from 75 thousand (K) hours of 64 million anonymized and transcribed single-talker English utterances, namely 75K dataset. The pre-training is based on the randomly simulated multi-talker audio segments from 75K data. For the binarized pattern training process, we are training our N-gram model from AMI meeting corpus \cite{carletta2005ami}, which consists of around 100 hours of meeting recordings, with each recording containing three to five speakers. For the evaluation, we evaluate our model performance in word error rate (WER) on AMI, ICSI \cite{janin2003icsi} and an internal Microsoft meeting dataset (MS).

Table \ref{tab:dataset} shows the dataset used in our experiments.
As public meeting evaluation sets, 
we used the AMI meeting corpus~\cite{carletta2005ami} and the ICSI meeting corpus~\cite{janin2003icsi}.
For the AMI corpus,
we used the first channel of the microphone array signals, also known as
the single distant microphone (SDM) signal.
For the ICSI corpus, we used the SDM signal 
from the D2 microphone. % (``sdm4'' in Kaldi toolkit \cite{}).
Both corpora were pre-processed and split into training, development,
and evaluation sets by using
the Kaldi toolkit~\cite{povey2011kaldi}.
In addition to these public corpora,
we also used 33 internal meeting recordings based on an SDM, 
denoted as MS$^{\rm mtg}$.
% for the evaluation, where the recording was made by a single distant microphone.
For all datasets, 
we applied causal
logarithmic-loop-based automatic gain control to normalize
the significant volume differences among different recordings.
Evaluation was conducted based on the utterance-group segmentation~\cite{kanda2021large},
and speaker-agnostic WER was calculated by using
 the multiple dimension Levenshtein edit distance~\cite{fiscus2006multiple,kanda2022vararray}.

% We trained 30-gram 
% As an instance of the statistical language model to represent the speech overlapping pattern,
% we trained 30-gram by using the AMI training data

For our proposed multi-talker audio simulation, 
the AMI training data with official word-level time stamps
was used to train the SLM, where we trained N-gram with $N=30$  
based on the NLTK toolkit \cite{bird2006nltk}.
For the source data ($S_0$) of the multi-talker audio simulation, we used 
75 thousand (K) hours of 64 million (M) 
anonymized and transcribed single-talker English utterances,
denoted as MS$^{asr}$ \cite{kanda2021large}.
 MS$^{asr}$ consists of 
audio signals 
from various domains, such as dictation and voice commands,
and each audio was
 supposed to contain single-talker speech.
 However, we found it occasionally
 contained untranscribed background human speech,
which broke the assumption for the multi-talker audio simulation.
 Therefore, we filtered out the audio sample
 that potentially contains more than two speaker audio.
 To detect such audio samples,
 we applied serialized output training-based multi-talker ASR 
 pre-trained by MS$^{\rm asr}$ and fine-tuned by AMI \cite{kanda2021large}
 for all audio samples in MS$^{\rm asr}$ with a beam size of 1.
 With this procedure, transcriptions of more than one speaker
 were generated for 14M utterances out of 64M utterances,
 which were 
 excluded in the multi-talker audio simulation.
 The effect of this data filtering is examined in Section \ref{subsec:basic}.

\begin{table}[t]
    % \vspace{-3mm}
    \caption{WERs (\%) of t-SOT TT18 (0.16-sec latency) trained with different algorithms for multi-talker audio simulation. 10K steps of fine-tuning was performed from the pre-trained t-SOT model.}
    %Fine-tuning was conducted for 10K steps with 8 GPUs where each GPU node consumed 24K frames of training samples for each training step. All evaluation data were SDM signals.}
    \label{tab:table1}
    \ra{0.9}
    \tabcolsep = 1.4mm
    \footnotesize
    \centering
    \begin{tabular}{cc|cc|cc|c}
    \toprule
    \multicolumn{2}{c|}{Fine-tuning data} & 
    \multicolumn{2}{c|}{AMI} & 
    \multicolumn{2}{c|}{ICSI} & 
    \multirow{2}{*}{MS$^{\rm mtg}$}  \\
    
     \multirow{1}{*}{Algorithm} & 
     \multirow{1}{*}{Filt.} & 
     \multirow{1}{*}{dev (1-spk / m-spk)} & 
     \multirow{1}{*}{eval} & 
     \multirow{1}{*}{dev} & 
     \multirow{1}{*}{eval} \\
    \midrule
    
 -  & -   & 35.5 (\textbf{24.4} / 42.5) & 39.2 & 32.2 & 33.3 & 27.3 \\
Rand ($K=2)$  & - &  35.9 (25.6 / 42.4) &	40.4 &	33.8 &	34.0 &	27.9 \\
Rand ($K=2)$  & $\surd$  &  31.6 (24.5 / 36.0) &	37.3 &	29.9 &	27.8 &	25.3 \\
Rand ($K=5)$  & $\surd$  &  31.2 (25.2 / 34.9) &	36.6 &	29.9 &	27.3 &	24.8 \\
Rand ($K=8)$  & $\surd$  & 31.1 (25.0 /	34.8) &	36.8 &	29.9 &	27.1 &	25.0 \\
    \midrule
 Word  & $\surd$ &  \textbf{30.6} (25.5 / \textbf{33.7}) &	36.1 &	\textbf{29.6} &	\textbf{26.5} &	25.0 \\
 Time ($d=0.10$)    & $\surd$ & 	31.4 (26.5 / 34.4)  &	36.0 &	30.2 &	27.4 &	24.7  \\
 Time ($d=0.25$) &    $\surd$ & 30.8 (26.0 / 33.8) &	\textbf{35.5} &	29.8 &	26.7 &	\textbf{24.5}  \\
 Time ($d=0.50$) &    $\surd$ & 32.6 (26.6 / 36.4) & 37.3 & 30.8 & 28.4 & 25.6  \\
    \bottomrule
    \end{tabular}
    \vspace{-5mm}
\end{table}

\subsection{Multi-talker ASR configuration}
\vspace{-.5em}
% \subsection{Multi-talker ASR configuration}
% \vspace{-.5em}

As an instance of multi-talker ASR, we trained t-SOT based 
 transformer transducer (TT) \cite{zhang2020transformer}
with chunk-wise look-ahead \cite{chen2021developing},
where
the algorithmic latency
of the model can be controlled
based on the chunk size of the attention mask.
We trained TT models
with 18 and 36 layers of transformer encoders,  which
were denoted by TT18 and TT36, respectively.
Each transformer block consisted of
a 512-dim multi-head attention with 8 heads 
and a 2048-dim point-wise feed-forward layer. 
The prediction network consisted of 
two layers of 1024-dim long short-term memory.
4,000 word pieces plus blank and $\langle \mathrm{cc}\rangle$ tokens
were used as the recognition units.
We used 80-dim log mel-filterbank extracted for every 10 msec.

All models were first pre-trained by using the 
multi-talker simulation data based on MS$^{\rm asr}$
with
the random simulation algorithm with $K=2$.
We performed 425K training steps with
 32 GPUs, with each GPU processing a mini-batch of 24K frames.
A linear decay learning rate schedule with a peak learning rate 
 of 1.5e-3 
 after 25K warm-up iterations were used.
 After the pre-training, 
 the model was further fine-tuned by using the simulation data
 based on  MS$^{\rm asr}$ and/or AMI \& ICSI training data.
 We performed various fine-tuning configurations, which will be presented in the 
 next section.

\begin{table}[t]
    % \vspace{-3mm}
    \caption{WERs (\%) of t-SOT TT18 (0.16-sec latency) with different combinations of multi-talker simulation algorithms. 20K steps of fine-tuning was performed from the pre-trained t-SOT model.} 
    % Fine-tuning was conducted for 20K steps with 8GPUs where each GPU node consumed 24K frames of training samples for each training step.}
    \label{tab:table2}
    \ra{0.9}
    \tabcolsep = 1.6mm
    \footnotesize
    \centering
    \begin{tabular}{ccc|cc|cc|c}
    \toprule
    % \multirow{3}{*}{Simulation Method} &
    \multicolumn{3}{c|}{\shortstack{Fine-tuning data}} &
    \multicolumn{2}{c|}{AMI} &
    \multicolumn{2}{c|}{ICSI} &
    \multirow{2}{*}{MS$^{\rm mtg}$}  \\
    
    \multirow{1}{*}{rand} &
    \multirow{1}{*}{word} &
    \multirow{1}{*}{time} &
    \multirow{1}{*}{dev (1-spk / m-spk)} & 
     \multirow{1}{*}{eval} &
    \multirow{1}{*}{dev} &
    \multirow{1}{*}{eval} \\
    \midrule
     1.0  &  -   & - &	31.0 (25.1 / 34.6) &	36.4 &	29.6 &	26.9 &	24.8 \\
     -  & 1.0 & -   & 31.0 (26.3 / 34.0) & 	36.2 &	29.6 &	26.4 &	25.0 \\
     -  &  -   & 1.0  &  31.1 (26.6 / 33.9) &	35.9 &	29.8 &	26.7 &	24.6 \\
    \midrule
     0.8 & 0.2 & - & 30.5 (24.9 / 33.9) &	35.9 &	29.1 &	26.1 &	24.5 \\
     0.5 & 0.5 & - &	30.3 (24.8 / 33.7)  &	35.7 &	29.3 &	25.8 &	24.3 \\
     0.2 & 0.8 & - & 30.7 (25.5 / 33.9) &	35.7 &	29.5 &	26.1 &	24.8 \\
    \midrule
     0.8 &  -  & 0.2 & 30.5 (25.0 / 34.0) &	35.8 &	29.2 &	26.5 &	24.3 \\
     0.5 &  -  & 0.5 & 30.5 (25.0 / 33.9)  &	36.0 &	29.3 &	26.2 &	24.4 \\
     0.2 &  -  & 0.8 & 30.2  (25.0 / 33.4) &	35.4 &	28.8 &	25.9 &	\textbf{24.0} \\
    \midrule 
     0.1 & 0.1 & 0.8 &	30.3 (25.4 / 33.3) &	\textbf{35.3} &	29.0 &	25.9 &	24.1 \\
     0.2 & 0.2 & 0.6  &	\textbf{30.0} (25.2 / \textbf{33.1}) &	\textbf{35.3} &	28.9 &	25.5 &	\textbf{24.0} \\
      0.3 & 0.3 & 0.4  &	\textbf{30.0} (25.0 / 33.2) &	\textbf{35.3} &	\textbf{28.7} &	\textbf{25.4} &	\textbf{24.0} \\
      0.4 & 0.4 & 0.2  &	30.3 (\textbf{24.8} / 33.8) &	35.7 &	28.8 &	25.8 &	24.2 \\

    \bottomrule
    \end{tabular}
    \vspace{-5mm}
\end{table}

\subsection{Experimental results and analysis}
\label{subsec:exp}
\vspace{-.5em}

\subsubsection{Comparison of simulation algorithms}
\label{subsec:basic}
\vspace{-.5em}

The WERs with different simulation algorithms are summarized in Table \ref{tab:table1}.
In this experiment, we conducted 10K steps of fine-tuning with 8 GPUs where each GPU node consumed 24K frames of training samples for each training step.
A linear decay learning rate schedule starting from a learning rate of 1.5e-4 was used.
The 1st and 2nd rows are the results of random simulation with $K=2$ for 
both pre-training and fine-tuning.
Because the data and algorithm used for the pre-training and fine-tuning are the same,
we didn't observe any improvement by additional fine-tuning.
Then, we applied the data filtering explained in Section \ref{subsec:data} 
(3rd row).
We observed significant WER improvements for all evaluation sets, confirming the
effectiveness of the data filtering.
We also evaluated the effect of different $K$ as shown in the 3rd to 5th rows,
where we observed $K=5$
and $K=8$ provides better results than $K=2$.

Then, we evaluated the proposed simulation algorithms, whose results 
are shown in the bottom four rows.
We observed that the best results for AMI-dev, ICSI-dev and ICSI-eval were obtained
by the proposed method with word-based discretization while
the best results for AMI-eval and MS$^{\rm asr}$ were obtained
by the proposed method with time-based discretization with $d=0.25$.
We also noticed that 
both of the proposed methods improved the WER of multi-speaker (m-spk) regions,
while had a degradation of the performance for the single-speaker (1-spk) regions.
We speculate this was caused because we short-segmented the sample in $S_0$
when we created the pool 
in the simulation process (see footnote 1), 
which introduced an unnatural onset / offset of the audio.

\begin{table}[t]
    \caption{WERs (\%) of t-SOT TT18 (0.16-sec latency) and t-SOT TT36 (2.56-sec latency) with different combinations of real training data 
    (AMI and ICSI) and proposed simulation training data.}
    %fine-tuning on a mixture of corpora. For the first row, the fine-tuning is 2,500 steps without gradient accumulation and with batch size as 12,000. For the rest, the fine-tuning is 20,000 steps with gradient accumulation as 2 and batch size as 12,000. (need to emphasize streaming)}
    \label{tab:table3}
    % \hspace*{-5mm}
    \ra{0.9}
    \tabcolsep = 1.2mm
    \footnotesize
    \centering
    \begin{tabular}{c|cc|cc|cc|c}
    \toprule
    % \multirow{2}{*}{Fine-tune Corpus} & 
    \multirow{2}{*}{Model} & 
    \multicolumn{2}{c|}{Fine-tuning data} & 
    \multicolumn{2}{c|}{AMI} & 
    \multicolumn{2}{c|}{ICSI} & 
    \multirow{2}{*}{MS$^{\rm mtg}$}  \\
   
    & 
    \multirow{1}{*}{AMI+ICSI} & 
    \multirow{1}{*}{MS$^{\rm asr}$-sim} & 
    \multirow{1}{*}{dev} & 
    \multirow{1}{*}{eval} & 
    \multirow{1}{*}{dev} & 
    \multirow{1}{*}{eval} \\
    \midrule
   \multirow{5}{*}{t-SOT TT18} &  1.0 &  -   & 21.9 & 25.7 &	20.3 & 17.9 & 24.9 \\
                               &  0.5 &  0.5  & 21.6 & 25.4 &	19.5 &	17.1 &	24.6	\\
                               &  0.2 &  0.8 &	21.8 &	25.5 &	19.9 &	17.3 &	23.4 \\
                               &  0.1 &  0.9 &	22.4 &	26.1 &	21.0 &	18.2 &	23.1 \\
                               &  - & 1.0  &	30.0 &	35.3 &	28.7 &	25.4 &	24.0 \\
   \midrule 
   \multirow{2}{*}{t-SOT TT36} &  1.0 &  -   & 16.9 & {\bf 19.7} &15.6 &14.0 &19.9 \\
                               &  0.2 &  0.8 & {\bf 16.8} & {\bf 19.7} &	{\bf 15.3} & {\bf 13.8}	 & {\bf 18.9} \\
    \bottomrule
    \end{tabular}
    \vspace{-5mm}
\end{table}

% previous table 3
% \begin{table*}[]
%     \centering
%     \begin{tabular}{c|c|c|cccc|cc|c}
%     \toprule
%     \multirow{3}{*}{Fine-tune Corpus} & \multirow{3}{*}{\shortstack{Corpus \\ Weights}} & \multirow{3}{*}{\shortstack{Mixture \\ Probabilities}} & \multicolumn{4}{c|}{AMI} & \multicolumn{2}{c|}{ICSI} & \multirow{3}{*}{DF33}  \\
%     & & & \multirow{2}{*}{dev} & \multicolumn{2}{c}{dev breakdown} & \multirow{2}{*}{eval} & \multirow{2}{*}{dev} & \multirow{2}{*}{eval} \\
%     & & & & 1-spk & m-spk & & & \\ 
%     \midrule
%     AMI    & 1.0 / \  0  & - &	18.0 & 14.1 & 20.4 &	21.3 &	23.6 &	20.7 &	24.3 \\
%     AMI\_ICSI    & 1.0 / \  0  & - & 17.8 & 14.2 &	20.1 &	20.8 &	15.9 &	14.1 &	23.2 \\
%     % AMI \& filt75K    & 0.5 / \  0.5  &	0.2 / 0.2 / 0.6 &	16.8 &	13.5 &	18.9 &	19.8 &	21.1 &	18.0 &	20.0 \\
%     % AMI\_ICSI \& filt75K    & 0.5 / \  0.5  & 0.2 / 0.2 / 0.6 &	16.6 &	13.5 &	18.6 &	19.7 &	15.1 &	13.5 &	19.6 \\
%     AMI \& filt75K    & 0.5 / \  0.5  &	0.3 / 0.3 / 0.4 &	16.8 &	13.5 &	18.8 &	19.8 &	20.8 &	18.0 & 20.0 \\
%     AMI\_ICSI \& filt75K    & 0.5 / \  0.5  & 0.3 / 0.3 / 0.4 &	16.6 &	13.4 &	18.6 &	19.6 &	15.4 &	13.3 & 19.8 \\

%     \bottomrule
%     \end{tabular}
%     \caption{WER (\%) of fine-tuning on a mixture of corpus for 20,000 steps. By default, filt75K is trained with mixed simulation as the simulation method, and the numbers in the column of Mixture Probabilities are in the order of random, word-based N-gram and time-based N-gram simulation respectively.}
%     \label{tab:table3}
% \end{table*}

\vspace{-.5em}
\subsubsection{Combination of multiple simulation algorithms}
\vspace{-.5em}

Given the observation about the trade-off between 1-spk and m-spk performance, we investigated 
the combination of the data from different simulation algorithms. 
The results are shown in Table \ref{tab:table2}. 
In this experiment, 
we increased the fine-tuning steps from 10K to 20K to make sure sufficient data were
generated
from each multi-talker audio simulation algorithm.
We used $K=5$ for the random simulation, and $d=0.25$ for
the time-based discretization.
From the table, we can observe that the trade-off between 1-spk and m-spk regions were
effectively resolved by mixing different types of multi-talker simulation data sets.
The best results for all evaluation sets were 
obtained when we combined all three multi-talker audio simulation algorithms
with the mixture ratio of 0.3, 0.3, 0.4 for the random simulation,
the proposed simulation with word-based discretization,
and the proposed simulation with time-based discretization, respectively. 

% The first part shows the performance of individual simulation methods. Note that here we increase the number of steps from 10K to 20K to make sure that N-gram simulation has a comparable amount of data consumption to previous experiments, so we can see that the three rows in the first part are slightly different from the corresponding WER in Table \ref{tab:table1}. But since they are overall at the same level, we can conclude that increasing number of steps from 10K to 20K will not make much difference to simulation methods without any mixing. 

% The rest of the table shows the mixed simulation results with different combinations of mixture probabilities of the three methods. From the last group, we can observe that with the mixture probabilities as $0.3$, $0.3$, and $0.4$ respectively, the WER can achieve the best number for all the 5 evaluation datasets, which indicates that the mixed simulation can further improve the performance compared to each of the single simulation method. From the breakdown, the mixed simulation has improved both 1-spk and m-spk regions, which indicates that the mixture is balancing the trade-off as expected.

\vspace{-.5em}
\subsubsection{Combination of real and simulated multi-talker audio}
\vspace{-.5em}

Finally, we also evaluated the combination of real meeting data (AMI and ICSI training data)
and the proposed simulation data. For the simulation data,
we used the best combination of three simulation algorithms as in the previous section,
and we performed 20K steps of fine-tuning.
Note that we performed only 2.5K steps of fine-tuning with 8 GPUs (12K frames of mini-batch per GPU) 
when the fine-tuning data
does not include the simulation data to avoid over-fitting phenomena.
As shown in the table, we observed the best results for all evaluation sets
when we mixed the real and simulation data with a 0.2 / 0.8 ratio.

\section{Conclusion}
\vspace{-.5em}
\label{sec:conclusion}

% In this paper, we propose to use N-gram to extract overlap patterns from real corpora and simulate statistically realistic conversations, and the simulated conversations can improve the performance of multi-talker ASR. We have also demonstrated that a mixture of different simulation methods and fine-tune corpora can further reduce the WER by balancing the trade-off between single-speaker and multi-speaker regions.

\vspace{-.5em}
This paper presented improved multi-talker audio simulation techniques for multi-talker ASR modeling.
We proposed two algorithms to represent 
the speech overlap patterns 
based on an SLM, 
which was then used to simulate multi-talker audio with realistic speech overlaps. 
In our experiments using multiple meeting evaluation sets, we demonstrated that multi-talker ASR models trained with the proposed method consistently showed improved WERs across multiple datasets.

% \section{Acknowledgement}
%\vfill
%\pagebreak

% \section{COPYRIGHT FORMS}
% \label{sec:copyright}

% \section{RELATION TO PRIOR WORK}
% \label{sec:prior}

\vfill\pagebreak

\bibliographystyle{IEEEbib}
\bibliography{refs}

\begin{thebibliography}{10}

\bibitem{seide2011conversational}
Frank Seide, Gang Li, and Dong Yu,
\newblock ``Conversational speech transcription using context-dependent deep
  neural networks,''
\newblock in {\em Proc. Interspeech}, 2011, pp. 437--440.

\bibitem{hinton2012deep}
Geoffrey Hinton, Li~Deng, Dong Yu, George~E Dahl, Abdel-rahman Mohamed, Navdeep
  Jaitly, Andrew Senior, Vincent Vanhoucke, Patrick Nguyen, Tara~N Sainath,
  et~al.,
\newblock ``Deep neural networks for acoustic modeling in speech recognition:
  The shared views of four research groups,''
\newblock {\em IEEE Signal processing magazine}, vol. 29, no. 6, pp. 82--97,
  2012.

\bibitem{qian2016very}
Yanmin Qian, Mengxiao Bi, Tian Tan, and Kai Yu,
\newblock ``Very deep convolutional neural networks for noise robust speech
  recognition,''
\newblock {\em IEEE/ACM Transactions on Audio, Speech, and Language
  Processing}, vol. 24, no. 12, pp. 2263--2276, 2016.

\bibitem{ccetin2006analysis}
{\"O}zg{\"u}r {\c{C}}etin and Elizabeth Shriberg,
\newblock ``Analysis of overlaps in meetings by dialog factors, hot spots,
  speakers, and collection site: Insights for automatic speech recognition,''
\newblock in {\em Proc. Interspeech}, 2006, pp. 293--296.

\bibitem{chen2020continuous}
Zhuo Chen, Takuya Yoshioka, Liang Lu, Tianyan Zhou, Zhong Meng, Yi~Luo, Jian
  Wu, and Jinyu Li,
\newblock ``Continuous speech separation: dataset and analysis,''
\newblock in {\em Proc. ICASSP}, 2020, pp. 7284--7288.

\bibitem{raj2020integration}
Desh Raj, Pavel Denisov, Zhuo Chen, Hakan Erdogan, Zili Huang, Maokui He,
  Shinji Watanabe, Jun Du, Takuya Yoshioka, Yi~Luo, et~al.,
\newblock ``Integration of speech separation, diarization, and recognition for
  multi-speaker meetings: System description, comparison, and analysis,''
\newblock in {\em Proc. SLT}, 2021, pp. 897--904.

\bibitem{yu2017recognizing}
Dong Yu, Xuankai Chang, and Yanmin Qian,
\newblock ``Recognizing multi-talker speech with permutation invariant
  training,''
\newblock {\em Proc. Interspeech}, pp. 2456--2460, 2017.

\bibitem{seki2018purely}
Hiroshi Seki, Takaaki Hori, Shinji Watanabe, Jonathan~Le Roux, and John~R
  Hershey,
\newblock ``A purely end-to-end system for multi-speaker speech recognition,''
\newblock {\em Proc. ACL}, 2018.

\bibitem{chang2019end}
Xuankai Chang, Yanmin Qian, Kai Yu, and Shinji Watanabe,
\newblock ``End-to-end monaural multi-speaker {ASR} system without
  pretraining,''
\newblock in {\em Proc. ICASSP}, 2019, pp. 6256--6260.

\bibitem{chang2019mimo}
Xuankai Chang, Wangyou Zhang, Yanmin Qian, Jonathan~Le Roux, and Shinji
  Watanabe,
\newblock ``{MIMO-SPEECH}: End-to-end multi-channel multi-speaker speech
  recognition,''
\newblock in {\em Proc. ASRU}, 2019, pp. 237--244.

\bibitem{tripathi2020end}
Anshuman Tripathi, Han Lu, and Hasim Sak,
\newblock ``End-to-end multi-talker overlapping speech recognition,''
\newblock in {\em Proc. ICASSP}. IEEE, 2020, pp. 6129--6133.

\bibitem{kanda2020serialized}
Naoyuki Kanda, Yashesh Gaur, Xiaofei Wang, Zhong Meng, and Takuya Yoshioka,
\newblock ``Serialized output training for end-to-end overlapped speech
  recognition,''
\newblock {\em Proc. Interspeech}, pp. 2797--2801, 2020.

\bibitem{lu2021streaming}
Liang Lu, Naoyuki Kanda, Jinyu Li, and Yifan Gong,
\newblock ``Streaming end-to-end multi-talker speech recognition,''
\newblock {\em IEEE Signal Processing Letters}, vol. 28, pp. 803--807, 2021.

\bibitem{sklyar2021streaming}
Ilya Sklyar, Anna Piunova, and Yulan Liu,
\newblock ``Streaming multi-speaker {ASR} with {RNN-T},''
\newblock in {\em Proc. ICASSP}, 2021, pp. 6903--6907.

\bibitem{kanda2022streaming}
Naoyuki Kanda, Jian Wu, Yu~Wu, Xiong Xiao, Zhong Meng, Xiaofei Wang, Yashesh
  Gaur, Zhuo Chen, Jinyu Li, and Takuya Yoshioka,
\newblock ``Streaming multi-talker {ASR} with token-level serialized output
  training,''
\newblock {\em Proc. Interspeech}, pp. 3774--3778, 2022.

\bibitem{kanda2022vararray}
Naoyuki Kanda, Jian Wu, Xiaofei Wang, Zhuo Chen, Jinyu Li, and Takuya Yoshioka,
\newblock ``{VarArray} meets {t-SOT}: Advancing the state of the art of
  streaming distant conversational speech recognition,''
\newblock {\em arXiv preprint arXiv:2209.04974}, 2022.

\bibitem{kanda2021large}
Naoyuki Kanda, Guoli Ye, Yu~Wu, Yashesh Gaur, Xiaofei Wang, Zhong Meng, Zhuo
  Chen, and Takuya Yoshioka,
\newblock ``Large-scale pre-training of end-to-end multi-talker {ASR} for
  meeting transcription with single distant microphone,''
\newblock {\em Proc. Interspeech}, pp. 3430--3434, 2021.

\bibitem{landini2022simulated}
Federico Landini, Alicia Lozano-Diez, Mireia Diez, and Luk{\'a}{\v{s}} Burget,
\newblock ``From simulated mixtures to simulated conversations as training data
  for end-to-end neural diarization,''
\newblock {\em Proc. Interspeech}, 2022.

\bibitem{yamashita2022improving}
Natsuo Yamashita, Shota Horiguchi, and Takeshi Homma,
\newblock ``Improving the naturalness of simulated conversations for end-to-end
  neural diarization,''
\newblock {\em Proc. Odyssey}, 2022.

\bibitem{li2022recent}
Jinyu Li,
\newblock ``Recent advances in end-to-end automatic speech recognition,''
\newblock {\em APSIPA Transactions on Signal and Information Processing}, vol.
  11, no. 1, 2022.

\bibitem{carletta2005ami}
Jean Carletta, Simone Ashby, Sebastien Bourban, Mike Flynn, Mael Guillemot,
  Thomas Hain, Jaroslav Kadlec, Vasilis Karaiskos, Wessel Kraaij, Melissa
  Kronenthal, et~al.,
\newblock ``The ami meeting corpus: A pre-announcement,''
\newblock in {\em International workshop on machine learning for multimodal
  interaction}. Springer, 2005, pp. 28--39.

\bibitem{janin2003icsi}
Adam Janin, Don Baron, Jane Edwards, Dan Ellis, David Gelbart, Nelson Morgan,
  Barbara Peskin, Thilo Pfau, Elizabeth Shriberg, Andreas Stolcke, et~al.,
\newblock ``The {ICSI} meeting corpus,''
\newblock in {\em Proc. ICASSP}. IEEE, 2003, vol.~1, pp. I--I.

\bibitem{povey2011kaldi}
Daniel Povey, Arnab Ghoshal, Gilles Boulianne, Lukas Burget, Ondrej Glembek,
  Nagendra Goel, Mirko Hannemann, Petr Motlicek, Yanmin Qian, Petr Schwarz,
  et~al.,
\newblock ``The {Kaldi} speech recognition toolkit,''
\newblock in {\em Proc. ASRU}, 2011.

\bibitem{fiscus2006multiple}
Jonathan~G Fiscus, Jerome Ajot, Nicolas Radde, and Christophe Laprun,
\newblock ``Multiple dimension {Levenshtein} edit distance calculations for
  evaluating automatic speech recognition systems during simultaneous speech,''
\newblock in {\em Proc. LREC}, 2006, pp. 803--808.

\bibitem{bird2006nltk}
Steven Bird,
\newblock ``{NLTK}: the natural language toolkit,''
\newblock in {\em Proceedings of the COLING/ACL 2006 Interactive Presentation
  Sessions}, 2006, pp. 69--72.

\bibitem{zhang2020transformer}
Qian Zhang, Han Lu, Hasim Sak, Anshuman Tripathi, Erik McDermott, Stephen Koo,
  and Shankar Kumar,
\newblock ``Transformer transducer: A streamable speech recognition model with
  transformer encoders and {RNN-T} loss,''
\newblock in {\em Proc. ICASSP}, 2020, pp. 7829--7833.

\bibitem{chen2021developing}
Xie Chen, Yu~Wu, Zhenghao Wang, Shujie Liu, and Jinyu Li,
\newblock ``Developing real-time streaming transformer transducer for speech
  recognition on large-scale dataset,''
\newblock in {\em Proc. ICASSP}, 2021, pp. 5904--5908.

\end{thebibliography}

\end{document}